%% file: nodeprop.tex
\documentclass[submission,copyright,creativecommons]{eptcs}
\providecommand{\event}{ICLP 2022} 
\usepackage{underscore}           

\usepackage{amssymb}
\usepackage{hyperref}
\usepackage{booktabs}
\usepackage[english]{babel}
\input pheader22.tex

\begin{document}

\title{ 
A Gaze into the Internal Logic of Graph Neural Networks, with Logic
}
\author{Paul Tarau
\institute{University of North Texas
}
\email{paul.tarau@unt.edu}
}

\def \authorrunning{Paul Tarau}
\def\titlerunning{A Gaze into the Internal Logic of Graph Neural Networks, with Logic}

\date{}
\maketitle

\begin{abstract}
Graph Neural Networks share with Logic Programming  several key relational inference mechanisms. The datasets on which they are trained and evaluated can be seen as database facts containing ground terms. This makes possible modeling their inference mechanisms with equivalent logic programs, to better understand not just how they propagate information between the entities involved in the machine learning process but also to infer limits on what can be learned from a given dataset and how well that might generalize to unseen test data. 

This leads us to the key idea of this paper: modeling with the help of a logic program the information flows involved in learning to infer from the link structure of a graph and the information content of its nodes properties of new nodes, given their known connections to nodes with possibly similar properties. The problem is known as  {\em graph node property prediction} and our approach will consist in emulating with help of a Prolog program the key information propagation steps of a Graph Neural Network's training and inference stages.

We test our a approach on the {\em ogbn-arxiv} node property inference benchmark. To infer class labels for nodes representing papers in a citation network, we distill the dependency trees of the text associated to each  node into directed acyclic  graphs that we encode as  ground Prolog terms. 
Together with the set of their references to other papers, they become facts in a  database on which we reason with help of a Prolog program that mimics the information propagation in graph neural networks predicting  node properties. In the process, we invent  ground term similarity relations that help infer labels in the test set by propagating node properties from  similar nodes in the training set and we evaluate their effectiveness in comparison with that of the graph's link structure. 
Finally, we implement explanation generators that unveil performance upper bounds inherent to the dataset.

As a practical outcome, we  obtain a logic program, that, when seen as  machine learning algorithm,
performs close to the state of the art on the node property prediction benchmark.

\end{abstract}

{\bf Keyphrases}: {\em 
Logic Programming and Machine Learning,
Graph Neural Networks,
graph node property prediction,
ground term similarity relations,
dependency trees,
text graphs,
symbolic vs. neural AI.
}

\section{Introduction}

Graph Neural Networks (GNNs) are extensively overviewed in \cite{gnn}. They use the graph structure (that can be seen as a binary predicate describing the connections between nodes) as well as properties associated to nodes and edge, ground facts in a database in logic programming parlance. Modern GNNs propagate information through a neighbor aggregation mechanism. Historically, GNNs can be seen as generalizing the convolution operations applied to neighboring pixel properties in image processing systems in Convolutional Neural Networks (CNNs), an idea going back as far as \cite{lecun89}.
Our first observation is that this information propagation mechanism is also clearly similar to the (more general) relational reasoning steps of a logic program.

Secondly, the datasets on which GNNs are trained and evaluated can be seen as database facts containing ground terms. While in deep learning systems complex information relating attributes of atoms linked in a molecular structure or dependency-linked words in a natural language sentence is usually flattened into fixed size embeddings (e.g., the word2vec model of \cite{mikolov14}), we also have the ability to process it via symbolic similarity relations, an operation easily implementable in logic programming languages.

Together, these two observations bring us to modeling key elements of a GNN's inference mechanisms with a logic program emulating its learning and inference steps.

As a first benefit, this will help us to better understand how GNNs propagate information between the entities involved in the machine learning process. As a second benefit, the logic-based modeling of the flow of information between nodes will  help us infer limits on what can be learned from a given dataset and how well that might generalize to unseen test data. 

This leads us to the key idea of this paper: {\em modeling with a help of a logic program the information flows involved in learning properties of unknown nodes from the link structure of a graph and the information content of its known nodes, given their connections to known nodes with possibly similar properties}.
This is known as the  {\em node property prediction} problem, which requires, given a set of nodes in a graph and  properties labeling a subset of nodes, to infer the labels of the nodes that are missing them. 

When applied to a concrete dataset, our in-depth analysis will also result in a Prolog program that will reveal what information elements contribute the most to improved performance in terms of prediction and generalization potential.

Thus, a practical outcome will be to evaluate the performance and the limitations of a purely symbolic, logic based inference mechanism, against state of the art graph neural networks  that dominate several leaderboards for this problem. A secondary outcome is, by invoking reasoning algorithms in logic programming, an ability to explain the contribution of the heterogeneous data elements used as features by the GNNs and elicit the contribution of their node-to-node propagation mechanisms as means for building learned data models. 

Among the available node property prediction tasks, we have picked the {\tt ogbn-arxiv} citation network from the 
Open Graph Benchmark  (OGB) dataset \cite{ogb} which provides convenient download and conversion tools to several graph formats, as well as a rich 
leaderboard\footnote{\url{https://ogb.stanford.edu/docs/leader_nodeprop/\#ogbn-arxiv}}
 comparing performance  of as many as 47 (at this time) GNNs.

After conversion to a Prolog database of ground terms, that also incorporates terms obtained by parsing the abstracts of the {\tt arxiv} documents with a dependency parser, we devise a  generic reasoning algorithm that infers missing labels using both the  structure of the citation network and the Prolog terms representing the distilled content of the document abstracts. 

As part of the inference mechanism, we introduce several similarity relations between Prolog terms, most of them novel and with potential applications as a less structure-sensitive alternative to Prolog's unification. We also design a new, content-weighted voting algorithm that helps with  propagating to a node with an unknown label, the most likely to be accurate label from its neighbors or its content-wise similar peers. Finally, we elicit explanations about what works and why, and we discover some inherent limitations in the data, that set an upper-bound on the performance of neural and/or symbolic approaches.

As a note on the Prolog code used\footnote{\url{https://github.com/ptarau/StanzaGraphs/tree/main/logic}}, we rely on the declarative subset of the SWI-Prolog \cite{swi} ecosystem, including higher order predicates and aggregates, but refrain from direct use of side effects like asserted code or global variables. The code has been tested with SWI-Prolog 8.1.3 and relies on a download of the {\tt arxiv} dataset
 that we have converted to a Prolog ground fact database\footnote{\url{http://www.cse.unt.edu/~tarau/datasets/arxiv_all.pro.zip}} .

The rest of the paper is organized as follows.
Section \ref{data} describes the structure of a dataset used for node property prediction
and its logical view provided by its conversion to a set of ground Prolog facts.
Section \ref{algo} overviews our generic reasoning algorithm that propagates 
information  form neighboring nodes in the training set, relying both on the structure
of the  citation graph and the similarity relations between the ground terms attached to the nodes.
Section \ref{sims} introduces several general ground-term similarity measures.
Section \ref{perf} evaluates performance of different parameter choices, including selecting similarity measures and  fine-tuning the diversity of information propagation from neighbors and distant peers in the graph.
Section \ref{expl} describes  explanation generation algorithms that unveil salient
properties of the dataset, including upper bounds on what neural or symbolic programs can achieve
given its actual information content.
Section \ref{rel} overviews related work and section \ref{concl} concludes the paper.

\section{The Dataset : a Logical View}\label{data}



We will describe our the general ideas behind our approach while working with a concrete dataset, that allows us to experimentally evaluate them.

The OGB dataset collection \cite{ogb} features graph property, node property and link property prediction tasks.
We have picked from the collection's node property prediction subset\footnote{\url{https://ogb.stanford.edu/docs/nodeprop/}}
the {\tt ogbn-arxiv} dataset. It contains a directed graph with {\tt 169343  nodes} and  {\tt 1166243 edges}, representing the citation network between Computer Science papers at the {\tt arXiv} repository, with edges indicating citation-links to other papers in the same set. As an additional resource, raw texts of titles and abstracts are provided.
The task is to predict the 40 subject areas of the  papers (which are manually  labeled), thus a 40-class classification problem.
The {\em training set} contains papers published until 2017, {\em the validation set} contains papers published in 2018, and {\em the test set} contains those published since 2019.

After digesting the dataset, a Python-based converter\footnote{\url{https://github.com/ptarau/StanzaGraphs/blob/main/logic/to_prolog.py}} is used to generate a set of  ground Prolog facts.

\subsection{Structure of the Citation Network}

The   ground Prolog facts are stored in a ``universal relation'' predicate {\tt at/5}, describing what information is located at each node. Each fact contains the following items:

\BE
\I a node identifier form {\tt 0 to 169342}
\I a marker in the set \verb~{tr, va, te}~, indicating if the fact belongs to the training, validation or test set
\I a label indicating one of the 40 classes, from {\tt 0} to {\tt 39}
\I a Prolog term representing a Directed Acyclic Graph (DAG), derived by merging the dependency trees of each sentence of the raw text associated to each node
\I a  list of  each node's neighbors, containing the node identifiers of the cited papers associated to the node.
\EE

\subsection{Merging Dependency Trees into Text DAGs}

We use the Stanza\footnote{\url{https://stanfordnlp.github.io/stanza/}} graph dependency parser \cite{stanza} 
developed by the Stanford NLP group to extract dependency trees for each sentence of a text obtained by concatenating the title and the abstract associated to a node in the {\tt arxiv} citation network.
Our preference for using a dependency parser rather  than a constituency parser
is that dependency trees can be seen as aggregating content elements in their natural order (e.g., arguments below predicates corresponding to verbs and adjectives below nouns that they refine semantically).

In a way similar to \cite{tplp20}, we post-process the dependency trees. While in \cite{tplp20} the post-processing step was generating a document graph to be used via centrality algorithms similar to \cite{EMNLP:TR} to extract summaries and keyphrases, this time we distill them into DAGs by reversing their dependency links, eliminating possible cycles and filtering out the terms unlikely to be relevant for comparing the content of two texts. 
We will illustrate the complete process over a short paragraph comparing logic and functional programming: 

\BOX{\small
{\em Logic Programming and Functional Programming are declarative programming paradigms. }
Evaluation in Logic Programs is eager while for functional programs it may be lazy. 
As a common feature, unification is used for evaluation in Logic Programming and 
for type inference in Functional Programming.
}

\subsection{From Text DAGs to Ground Terms}

Given the isomorphism between a ground Prolog term, its tree representation and the DAG obtained by fusing together its shared subterms, we obtain the following Prolog term representing the paragraph:\\

\BOX{\small text\_term(paradigm(programming(logic, functional, inference(type)), declarative), eager(evaluation(programming(logic, functional, inference(type)), program(logic, functional)), lazy(program(logic, functional))), use(evaluation(programming(logic, functional, inference(type)), program(logic, functional)), feature(common), unification))
}\\

\section{The Label Inference Algorithm}\label{algo}

Approaching algorithmically a machine-learning problem requires a careful mapping of its traditional data splitting conventions to a mechanism separating what information is available to the algorithm and what ought to be hidden from it.
In a machine learning context, the nodes are divided into a training set, a validation set and a test test. The key assumption is that learning occurs by trying to predict labels in the validation set using information from the training set. To ensure that generalization occurs, only inference is applied to the test set, which must be carefully hidden from the learning algorithm. However, when one works  algorithmically on the same problem, with no machine learning involved, the training set and the validation set can be merged into an extended training set at the disposal of the inference algorithm, whose performance will then be evaluated against the carefully hidden test set.

While there are a wide variety of GNNs, as for instance those collected at by the Pytorch Geometric project \cite{pyg}, their shared focus originates from Convolutional Neural Networks \cite{cnn} (CNN). To ensure translation and scale invariance as well as higher-level observables from images, CNNs work by propagating features of neighboring pixels. As a generalization of this mechanism, GNNs  propagate information in graphs, between a node and its neighbors.

At a first glance, a citation network should be a DAG, but in practice cycles can form easily, for instance when two papers cite each other. Also, papers in the test set might cite other papers in the test set and that information needs to be disallowed. Citation information can also be absent or severely reduced when a paper cites few or no other papers in the  dataset on which the network is based. This requests the use of content elements in the nodes to complement information gleaned from the structure of graph. 
Full details of our code, ensuring that such properties hold, are hosted on github\footnote{\url{https://github.com/ptarau/StanzaGraphs/blob/main/logic/thinker.pro}}.

To infer the label associated to a node, the first thing to do is to analyze the labels of its references, when known. At the same time, to complement it or replace it when this information is not available, content similarity relations need to be used to propagate information from potentially any node in the graph, but ideally with some guidance from similarities between the labels themselves, if that can be inferred from the data.

Options for the label inference algorithm are guided by a predicate {\tt param/2} supporting
different combinations of algorithm components and parameters guiding its focus, accuracy and performance.

The predicate accuracy/1 computes the ratio between correct inferences and total facts in the test set, using the library predicate {\tt aggregate\_all/3}\footnote{\url{https://www.swi-prolog.org/pldoc/man?predicate=aggregate_all/3}}. As it is usual for machine learning programs, we prefix with {\tt Y} the label variables to be trained on and to be guessed in inference mode.
\begin{code}
accuracy(Acc):-
   test_size(Total), 
   aggregate_all(count,correct_label,Success),
   Acc is Success/Total.
\end{code}
\begin{code}
correct_label:-inferred_label(YtoGuess, YasGuessed), YtoGuess = YasGuessed.
\end{code}

The predicate {\tt inferred\_label} returns pairs  {\tt YtoGuess}, {\tt YasGuessed} and it is parameterized by limits on the number of nodes  to be used when exploring via a similarity relation neighbors or other peer nodes. 

For each node in the test set, provided by the predicate {\tt tester\_at/4}, we explore 3 alternative inference mechanisms, whose output is given to a weighted voting algorithm {\tt vote\_for\_best\_label} that picks the inferred label.

The predicate {\tt neighbor\_data/5} receives {\tt MyTextTerm} denoting the Prolog term representing the content of a node to be computed based on a similarity weight for each of its selected neighbors. 
\begin{code}
inferred_label(YtoGuess, YasGuessed):-
   param(max_neighbor_nodes,MaxNodes),
   param(neighbor_kind,NK), 
   most_freq_class(FreqClass),
   select_diverse_peers(Peers), 
   tester_at(N,YtoGuess,MyTextTerm,Neighbors),
   ( NK\=none, at_most_n_sols(MaxNodes,YW,
      neighbor_data(NK,Peers,MyTextTerm,Neighbors,YW),YsAndWeights)->true
   ; peer_data(Peers,MyTextTerm,YsAndWeights)->true
   ; YsAndWeights=[FreqClass-1.0] 
   ),vote_for_best_label(YsAndWeights,YasGuessed).
\end{code}
If similar enough neighbors are found, no other algorithm is invoked and the results are passed to the voting predicate.
If no such neighbors are found, the predicate {\tt peer\_data/3} will try  to extract weighted {\tt Y} values  from nodes with similar textual descriptions from the training set. Should this also fail, the label of the most popular category in the training set is assigned as a default label. 

The astute reader might notice that here we are, in fact, emulating a GNN's message passing step aggregating information from neighboring nodes, as well as from peer nodes similar but possibly located far away in the graph.

An important mechanism to ensure accuracy in {\tt peer\_data/3} is {\em diversity}. It is enforced by comparing against nodes in the training set evenly distributed among the available labels. The set of these preselected nodes is passed as the variable {\tt Peers}. Ensuring that enough distinct peers (by default 4) are available helps with improved performance as it will be discussed in section  \ref{perf}.

Note also  that  {\tt neighbor\_data/5} is  parameterized (see variable {\tt NK}) offering a choice to enforce or not diversity of the labels of the neighbors.   If one wants to test out performance without using information about the graph's structure, one can disable this choice with option {\tt none}.

The predicate {\tt vote\_for\_best\_label/3} implements our weighted voting mechanism. After summing up the weight of each label of the same kind, it picks the one with the highest total weight using the library predicate {\tt max\_member}. Grouping by labels is achieved with help of {\tt keygroups/2} relying itself on the built-in {\tt keysort/2} and the library predicate {\tt group\_pairs\_by\_key/2}.

\begin{code}
vote_for_best_label(YsAndWeights,YasGuessed):-
   keygroups(YsAndWeights,YsAndWeightss),
   maplist(sum_up,YsAndWeightss,WeightAndYs),
   max_member(_-YasGuessed,WeightAndYs).
 \end{code}
 
 \begin{code}  
 keygroups(Ps,KXs):-
   keysort(Ps,Ss),
   group_pairs_by_key(Ss,KXs).
\end{code}

Note that our (weighted) voting mechanism implemented by the predicate {\tt vote\_for\_best\_label} is in fact an approximation of the message-passing information propagation mechanism \cite{gnn} giving the peers of a node a say on the node's decision to pick a given label in a classification task.
 
Next, we will expand on some details of the similarity measures that provide weights to our voting algorithm inferring for each node the most likely to be accurate label.

\section{Ground Term Similarity Measures}\label{sims}

The code covering this section is in file {\tt sim.pro} at github\footnote{\url{https://github.com/ptarau/StanzaGraphs/blob/main/logic/sims.pro}}.
We have approached computation of similarity relations between Prolog terms with a broader application scope in mind. 
One limitation of unification as a pattern matching algorithm against a ground term database is that matching exact argument numbers and positions might be too strict for possibly noisy data as it would be the case for the Prolog terms synthesized from the dependency trees described in section \ref{data}. Unification would also make similarity relations trivial, as  it collapses to structural identity in the case of two ground terms.

The predicate {\tt shared\_path/3} extracts segments of paths in the trees corresponding to two terms, assuming that the order of the arguments is irrelevant and that distinct functor symbols at the same level can be skipped. We call them ``pathlets'' by analogy to the graphlets used for pattern matching on graphs \cite{graphlets}.
\begin{code}
shared_path(S,T,Path):-distinct(Path,shared_path(S,T,Path,[])).

shared_path(S,T)-->{atomic(S),atomic(T)},!,emit_atom(S,T).
shared_path(S,T)-->{atomic(S),functor(T,F,_)},!,emit_atom(S,F).
shared_path(S,T)-->{atomic(T),functor(S,F,_)},!,emit_atom(T,F).
shared_path(S,T)-->
  {functor(S,F,_),functor(T,G,_)},
  emit_atom(F,G),
  {arg(_I,S,X),arg(_J,T,Y)},
  shared_path(X,Y).

emit_atom(S,S)-->!,[S].
emit_atom(_,_)-->[].
\end{code}

\BX Computing shared paths between two terms
\begin{codex}
?- shared_path(g(f(a(1)),b(1),c),h(f(a(2)),b(1),c),R).
R = [f, a] ; R = [b, 1] ; R = [c].
\end{codex}
\EX
From this predicate, we derive a similarity measure {\tt shared\_path\_similarity/3} by wrapping the call to {\tt shared\_path} into an {\tt aggregate\_all} predicate summing up the length of these shared ``pathlets''. 

Alternatively, {\tt forest\_path\_similarity} can first split a few levels at the top of the tree into forests, thus avoiding some boiler-plate shared content close to the root.

The predicate {\tt termlet\_similarity}  is derived from summing up sizes of subterms up to a give size, shared by the two terms.
\begin{code}
termlet_similarity(A,B,Sim):-
   param(max_termlet_size,MaxTS),
   aggregate_all(sum(Sim),sharing_count(MaxTS,A,B,Sim),Sim).
\end{code}

It calls  {\tt sharing\_count} which sums up the sizes of all shared subterms. The SWI-Prolog built-in {\tt sub\_term} is used to backtrack over subterms of the {\tt A}, after which occurrences of these subterms in {\tt B} are counted.
\begin{code}
sharing_count(MaxTS,A,B,Res):-
  sub_term(T,A),term_size(T,Size0),
  Size is 1+Size0,Size=<MaxTS,
  occurrences_of_term(T,B,Count),
  Count>0,Res is Size*Count.
\end{code}

We have also defined two similarity measures, based on the Jaccard  index \cite{jaccard} between sets of nodes and sets of edges in the trees associated to the ground Prolog terms. We remind the reader that the Jaccard similarity $J_s$ is computed by dividing the size of the intersection of two sets with the size of their union:\\

$ J_s(A,B) = {|{A \cap B}| \over |{A \cup B}|} = {|{A \cap B}| \over |A| + |B| - |{A \cap B}|} $\\

\noindent Its corresponding distance $J_d$, defining a metric space structure on finite sets is:\\

$ J_d(A,B) = 1-J_s(A,B) $\\

Finally, as a way to enable an unweighted (``democratic'') vote among neighbors of a node proposing their labels as candidates, we define:
\begin{code}
mock_similarity(_A,_B,1.0).
\end{code}
Thus, based on this definition,  every term is similar to every other term.

\section{Evaluating Accuracy}\label{perf}

We have run as a baseline accuracy the Pytorch Geometric-based GNN implementation in the OGB repository\footnote{\url{https://github.com/snap-stanford/ogb/blob/master/examples/nodeproppred/arxiv/gnn.py}} and we have also compared our purely symbolic algorithm with the best GNN-based result on OGB arxiv leaderboard. We have used several parameter combinations, with focus on assessing the impact of the graph structure vs. the impact of the content associated to the nodes. We have summarized in Table \ref{eval} the accuracy of our algorithm parameterized by the similarity relation used, and in Table \ref{geval} two GNN-produced results. 

The first line in Table \ref{eval} contains our results using both the structure of the graph and its node content similarity. 
The second line collects ``node content similarity only'' results obtained by disabling in the predicate {\tt inferred\_label/2} the call to {\tt neighbor\_data/5}. 

With respect to the GNNs, our best result (with {\tt jaccard\_node\_similarity}) is less than 4 points below the OGB GNN and less than 6 points below the best result on the leaderboard, but, surprisingly, still ahead of the first attempt (May 2020) by the OGB team at 0.5765 and a few other GNN-based results. 

Could we have done better by writing a cleverer program? A hint that this is unlikely, comes from the embeddings used by the GNNs. The textual content, as used by the GNNs on the arxiv benchmark is transformed via embeddings to 
128-long floating point vectors that are known to better capture contextual information about content similarity than our discrete similarity measures on the content words and their local dependencies, as they are trained on very large corpora.
As the second line in Table \ref{eval} reveals, the impact of the node similarities is significantly lower than that of the link structure of the citation network. Embeddings also have meaningful distance preservation and algebraic properties when subject to aggregation operations like dot product, addition and subtraction, given that they operate in the metric space $\mathbb{R}^N$.

The comparatively good performance of the {\tt jaccard\_node\_similarity} that does not directly use information about the structure of our trees can also be explained by the fact that its related Jaccard distance $J_d$ defines a metric space, a property that does not hold for the other similarity relations.

More surprising is the  unusually good performance of {\tt mock\_similarity}. By always returning $1$ independently of its arguments, this relation simply gives equal weights to each of the neighbors and peers of a node. Thus, the resulting  ``democratic'' voting mechanism  eliminates the impact of the significantly weaker content element while letting the graph's structure decide alone on the choice of the most likely label. While the two best similarity relations still outperform it, the weaker ones, as illustrated by the second row in Table \ref{eval}, show that weighting the votes based on a weak similarity relation can actually hinder performance.



\begin{table}[]
\centering
{
\begin{tabular}{|l|c|c|c|c|c|c|}
\cline{1-7}
\textbf{similarity measure used} & mock   & termlet & shared\_path & forest & jaccard\_edge & jaccard\_node   \\ \cline{1-7}
\textbf{with graph structure}    & 0.6719 & 0.6639  & 0.6780       & 0.6713 & 0.6624        & \textbf{0.6860}  \\ \cline{1-7}
\textbf{content similarity only} & 0.0586 & 0.0815  & 0.1558       & 0.1476 & 0.1282        & \textbf{0.2825}  \\ \cline{1-7}
\end{tabular}%
}
\vskip 0.5cm
\caption{
Performance impact of graph structure vs. similarity based node content}
\label{eval}
\end{table}

\begin{table}[]
\centering
{
\begin{tabular}{|l|c|c|}
\cline{1-3}
\textbf{state of the art GNN performance}  & {\color[HTML]{00009B} OGB GNN} & {\color[HTML]{00009B} Leaderboard GNN} \\ \cline{1-3}
\textbf{with graph structure and node content information}   &  {\color[HTML]{00009B} 0.7213}  & {\color[HTML]{00009B} \textbf{0.7431}} \\ \cline{1-3}
\end{tabular}%
}
\vskip 0.5cm
\caption{ GNN performance}
\label{geval}
\end{table}

The predicate {\tt param/2} defines our configuration parameters. We have used the same (relatively small) values across all similarity measures as execution time would be prohibitive otherwise for slower ones like {\tt termlet\_similarity}.

\BX
We run our tests for all similarities with the following parameters:
\begin{codex}
param(max_neighbor_nodes,100). 
param(max_peer_nodes,4). 
param(similarity, node_jaccard_similarity). 
\end{codex}
resulting  for the best performing {\tt jaccard\_node\_similarity} measure, in an accuracy of {\bf 0.6860} with a runtime of {\tt 108.50 seconds}.
\EX

The astute reader might notice at this point that our choice of similarity relations is a discrete, unlearn approximation of the implicit similarity relations learned via gradient descent in the backpropagation steps of a GNN. Therefore it is also somewhat surprising that we perform within a fairly close margin to the best GNNs on the leaderboard and outperform several other GNNs on it.

\section{The Explainer}\label{expl}

Explainability of AI-results \cite{explAI} is an important parameter not just for improving performance of, but also for helping the human stakeholders trust AI-applications. 
In a deep learning context its focus is on discovering human-understandable descriptions of the internal states of learned parameters.

In our logic programming context we will explore next explanations derived from our ``dissection'' of the structure of the dataset and the effectiveness of the algorithms used to infer the unknown node properties as well as the projected upper limits on performance, inherent to the information available in the dataset.

\subsection{The Impacts}

\subsubsection{Graph Structure vs. Node Content}

The results in Table \ref{eval} show that when we hide the connectivity information of the graph and use exclusively node content similarity, our best result is still only {\tt 0.2825}. In fact, before thinking about making sure that the random selection of peers is replaced by a diverse batch that contains several peers of in each class of the training set (4 peers by default) the results were below {\tt 0.10 for all similarity measures}. But can this be improved by offering a larger selection of peers?

\BX \label{forty}
When pushing a few steps further the number of peers selected for each label a significant increase is noticed, at the expense in this case of almost 3hours for running the test:

\begin{codex}
param(max_neighbor_nodes, 100).
param(max_peer_nodes, 40).
param(similarity, node_jaccard_similarity).
\end{codex}
The resulting accuracy of {\bf 0.4637} shows that {\em peer diversity} brings a significant increase of the effectiveness of our node content similarity measures, despite the absence of information about the structure of the network.
\EX

\subsubsection{Diversity}
This significant improvement in accuracy motivates increasing the number of diverse peer nodes when evaluating their impact in the presence of information about the graph's structure.

\BX When running with a much larger sample of {\tt 200} peers for each label:
\begin{codex}
param(max_neighbor_nodes, 400).
param(max_peer_nodes, 200).
param(similarity, node_jaccard_similarity).
\end{codex}
we can lift accuracy to {\bf 0.6930} with a runtime of {\tt 1694.20} seconds, but quickly reaching diminishing returns after that. Nevertheless, that  brings us less than 3 points below the OGB GNN result.
\EX

As the majority of the nodes make their label guesses based on their neighbors, the improvement is, as expected, not very large. This brings us to try out what happens if we enforce diversity also for neighbors.

\BX
Should we enforce diversity among a node's neighbors?
\begin{codex} 
param(max_neighbor_nodes, 100).
param(max_peer_nodes, 4).
param(neighbor_kind, diverse).
param(similarity, node_jaccard_similarity).
\end{codex}
Unfortunately we notice a drop in performance of accuracy from {\bf 0.6830} to {\bf 0.6607}. A similar drop happens also from {\bf 0.6780} to {\bf 0.6508} for {\tt shared\_path\_similarity}.

\EX

Thus, while diversity is good when relying on content similarity with possibly far-away peers in the network, it actually harms performance when enforced on the neighbors bringing in information about the structure of the graphs. The explanation has to do with the fact that the possibly uneven distribution of the cited papers in each node is actually an advantage. With  papers usually citing a majority of papers in their own (to be predicted) class, letting them vote for it is better than forcing them to pick evenly from all possible classes.


\subsection{The Limits}

We start with observing the clustering at the top for both the OGB leaderboard
\footnote{with as many as  {\bf 8}   results between {\bf 0.7402} and {\bf 0.7431} on the test set} and the parameter choices for our symbolic approach. Why are such different approaches hitting a wall around a small performance window?
This brings us to the need to discover inherent limits that would apply to both symbolic and neural approaches within their own performance margins, for which we will describe here a few predicates from the file {\tt explainer.pro}\footnote{\url{https://github.com/ptarau/StanzaGraphs/blob/main/logic/explainer.pro}}.

\subsubsection{How much we can learn from graph's structure}

We will first compute the size of total number of testable nodes:
\begin{code}
testables(Count):-aggregate_all(count,N,testable(N),Count).
testable(N):-tester_at(N,_Y,_T,_Ms).
\end{code}

Next we compute the number of ``guessable nodes'' for which it is possible, using exclusively the link structure of the graph, to predict their label. The main idea is to  check if, for a given node in the test set, there's at least one neighbor in the training set that has the same label as the one we need to correctly predict.

\begin{code}
network_guessables(Count):-
  aggregate_all(count,N,network_guessable(N),Count).
\end{code}

\begin{code}
network_guessable(N):-
  tester_at(N,Y,_T,Ms),
  once((member(M,Ms),trainer_at(M,Y,_,_Ms))).
\end{code}
This gives us the limits:
\begin{code}
network_limits:-
  do((
    network_guessable_ratio(Result),
    writeln((network_only->Result))
  )).
\end{code}
\begin{code}
network_guessable_ratio(R=GCount/TCount):-
  testables(TCount),
  network_guessables(GCount),
  R is GCount/TCount.
\end{code}

With this in mind, the predicate {\tt network\_limits/1} computes an upper limit of {\bf 0.8441} to what an oracle picking always the right label from a neighbor could achieve.

\BX Computing the upper limit of how much can be inferred from the network structure
\begin{codex}
?- network_limits.
network_only -> 0.8441 = 41028/48603
\end{codex}
\EX

\subsubsection{How much we can learn from the node content - in theory and in practice}

We have also written a few predicates in {\tt explainer.pro} meant to infer upper limits on how much can be predicted using exclusively  content similarities between nodes. In theory, given a minimal threshold associated to each similarity there's almost always a close enough other node.

\BX  Nodes having a similar peer above a given threshold for each measure.
\begin{codex}
node_jaccard_similarity-0.01->48603/48603=1
shared_path_similarity-1.0->48488/48603=0.9976
edge_jaccard_similarity-0.01->44529/48603=0.9161
forest_path_similarity-1.0->47298/48603=0.9731
\end{codex}
\EX
In practice this is not very useful however, as the large number  matches can easily move the vote towards a false positive. It also assumes a total of $O(N^2)$ similarity computations between a node and all the other nodes. Thus the most likely practical limit is somewhere close above  the {\bf 0.4637} value of Example \ref{forty} corresponding to a set of 40 * 40 = 1600 peer nodes with evenly distributed labels (assuming not more than a few hours of total computing time).

\subsection{Discussion}

An interesting question is raised by the possibility of using a deeper symbolic propagation mechanism, for instance, to compute neighbors in the transitive closure of the citation graph or just a few steps further. Our experiments have shown that even involving the neighbors of neighbors results in reduced performance, possibly because the amount of noise it brings in, even if the voting value of the second step neighbors is dampened (e.g.. reduced to half).

Once the doors for logical reasoning have been opened, a few other scenarios come to mind. One of them involves
hypothetical reasoning. From a given nodes's perspective, say node N, assuming that other nodes in the test set that are similar to N got label Y, then N  might also want to consider getting Y with some probability. This would be doable, assuming, as it is usual for typical ML datasets, that the test set is available as a block rather than one node at a time.
Also, similarity clusters can be discovered in the training set assuming it is given as a block. Then, such a community of nodes might propose a smaller set of dominant labels with outliers removed.

With regard to diversity, a likely improvement would be to follow the distribution of labels in the training set rather than an even one. In this case the number of nodes with label Y should mimic the frequency of Y in the training dataset, assuming that the label distribution in the test set is close to that in the training set.

\section{Related work}\label{rel}

Comparing neural and symbolic algorithms with focus on performance goes back as far as \cite{con_vs_sym} and \cite{neuro_vs_sym}. More recently, and \cite{deeplimits} offers a  comprehensive discussion of their limitations, as seen through the eyes of human domain experts. 

We refer to \cite{d2020neurosymbolic} for an overview of the fast-growing attempts to neuro-symbolic integration and to 
\cite{lamb2020graph} for an overview specialized to graph-neural networks, especially suitable for relational inference on unstructured data. In \cite{de2020statistical} a large number of neuro-symbolic systems are surveyed, along multiple features, among which  the expressiveness of the logic (ranging from propositional to first order predicate logic) and its probabilistic or deterministic nature. However, most of this work focuses on merging the two domains's best features in unified neuro-symbolic systems, rather than, as we do in this paper, explore via a logic program the structure of a typical deep learning dataset, to elicit information in an explainable way about the inference process itself and its model-independent limitations. Our voting algorithm shares ideas that have been used by ensemble classifiers in ML, for instance the highly effective Random Forests algorithm \cite{ranfor}, except that our algorithms are fully deterministic, with no random choices involved, with the advantage of being exactly replicable. 

The similarity relations that we have explored relate to the rich literature on graph kernels \cite{gkernels} as well as graph and tree edit distances \cite{tree_edit}. However, our focus in this paper is on similarities between DAGs expressed as ground Prolog terms, which can be seen as a follow-up to the symbolic and neural variants of the path-based indexing mechanism of ground Prolog terms in \cite{iclp21}. Another use of dependency trees in the context of a logic program is described in \cite{tplp20} with a similar technique of redirecting links to be used in text graphs. By contrast, our focus in this paper is on merging them into DAGs and ultimately into ground Prolog terms usable for similarity measures.

\section{Conclusions}\label{concl}
We have described a set of logic-based techniques for analyzing  datasets used in deep learning benchmarks together with a  symbolic solution that approximates via explainable reasoning steps key aspects of the information propagation happening in GNNs. While we have focussed on a concrete dataset ({\em ogbn-arxiv}) and a specific task (node property prediction) the general methodology, relying on exploring the analogy between information propagation in a GNNs and the equivalent reasoning steps implemented in Prolog is generalizable to other graph-based datasets, as well as link or graph property prediction tasks. While the same analysis, once expressed as a logic program, could be reimplemented as a Python or Julia module, the declarative simplicity of our code shows the effectiveness of a mature Prolog ecosystem as a competitive exploration tool revealing fine points about the internal logic of deep learning tools.

We have evaluated the impact and the limitations of the underlying graph structure as well as that of the content associated to the nodes. While we have observed that the graph structure dominates, we have also discovered that if possibly far away peer nodes are used instead of the assumed unavailable neighbors, diversity brought by ensuring that all labels are fairly represented can significantly improve results relying exclusively on content similarity measures. This gives an actionable hint on designing GNNs able to reach far away in the graph's link structure, guided by content-similarity measures.
Along these lines, porting our Prolog ground-term similarity measures can be beneficial as data-augmentation tools in a preprocessing step for GNNs. At the same time they are likely to have broader uses as more flexible  alternatives to unification, when data is represented in the form of ground Prolog terms. 

\bibliographystyle{eptcs}

\bibliography{theory,tarau,ml,nlp,proglang,biblio,ref}
\end{document}

%% file: pheader22.tex
\usepackage{graphicx}
\usepackage{mathptmx}
\usepackage{tikz-qtree}
\usepackage{filecontents}
\usepackage{csvsimple}
\usepackage{amsfonts,amsmath}
\usepackage{subfig} 
\usepackage{url}
\usepackage{verbatim}
\usepackage{color}
\usepackage{amsmath} 
\usepackage{proof}

\renewcommand{\phi}{\varphi}

\input prolog.tex

\definecolor{lgray}{gray}{0.95}
\definecolor{lblue}{rgb}{0.90,0.90,1.00}
\definecolor{lyellow}{rgb}{1.00,1.00,0.70}

\usepackage{listings}
\newtheorem{ex}{Example}

\lstnewenvironment{codex}
    {\lstset{}%
      \csname lst@SetFirstLabel\endcsname}
    {\csname lst@SaveFirstLabel\endcsname}
\newcommand{\BI}[0]{\begin{itemize}}
\newcommand{\EI}[0]{\end{itemize}}
\newcommand{\I}[0]{\item}
\newcommand{\BE}[0]{\begin{enumerate}}
\newcommand{\EE}[0]{\end{enumerate}}

\newcommand{\BX}[0]{\begin{ex}}
\newcommand{\EX}[0]{\end{ex}}
\newcommand{\BV}[0]{\begin{verbatim}}
\newcommand{\EV}[0]{\end{verbatim}}

\newcommand{\BF}[0]{\begin{filecontents*}{data.csv}}

\newcommand{\BQ}[0]{\color{blue}\begin{quote}}
\newcommand{\EQ}[0]{\end{quote}\color{black}}

\def \bscale1 {0.25}
\def \bscale {0.25}

\newcommand{\FIG}[4]{
\begin{figure}[htbp]
\centering
{\includegraphics[scale=#3]{#4}}
\caption{#2}
\label{#1}
\end{figure}
}

\newcommand{\BOX}[1]{
\noindent\fbox{%
    \parbox{\textwidth}{%
    #1
    }%
}}

\newcommand{\HFIGS}[7]{
\begin{figure}[htbp]
  \centering
  \subfloat[#3]{
    {\includegraphics[scale=#7]{#5}}}   
  \subfloat[#4]{
    {\includegraphics[scale=#7]{#6}}}
  \caption{#2}
  \label{#1}
\end{figure}
}

